\documentstyle[psfig]{mn}

\newif\ifAMStwofonts

\ifoldfss
  \ifCUPmtlplainloaded \else
    \NewTextAlphabet{textbfit} {cmbxti10} {}
    \NewTextAlphabet{textbfss} {cmssbx10} {}
    \NewMathAlphabet{mathbfit} {cmbxti10} {} 
    \NewMathAlphabet{mathbfss} {cmssbx10} {} 
  \fi
  \ifAMStwofonts
    \ifCUPmtlplainloaded \else
      \NewSymbolFont{upmath} {eurm10}
      \NewSymbolFont{AMSa} {msam10}
      \NewMathSymbol{\upi}     {0}{upmath}{19}
      \NewMathSymbol{\umu}     {0}{upmath}{16}
      \NewMathSymbol{\upartial}{0}{upmath}{40}
      \NewMathSymbol{\leqslant}{3}{AMSa}{36}
      \NewMathSymbol{\geqslant}{3}{AMSa}{3E}

       \let\le=\leqslant
       
    \fi
  \fi
\fi 

\ifnfssone
  \newmathalphabet{\mathit}
  \addtoversion{normal}{\mathit}{cmr}{m}{it}
  \addtoversion{bold}{\mathit}{cmr}{bx}{it}
  \newmathalphabet{\mathbfit} 
  \addtoversion{normal}{\mathbfit}{cmr}{bx}{it}
  \addtoversion{bold}{\mathbfit}{cmr}{bx}{it}
  \newmathalphabet{\mathbfss} 
  \addtoversion{normal}{\mathbfss}{cmss}{bx}{n}
  \addtoversion{bold}{\mathbfss}{cmss}{bx}{n}
  \ifAMStwofonts
    \ifCUPmtlplainloaded \else
      \UseAMStwoboldmath
      \makeatletter
      \new@mathgroup\upmath@group
      \define@mathgroup\mv@normal\upmath@group{eur}{m}{n}
      \define@mathgroup\mv@bold\upmath@group{eur}{b}{n}
      \edef\UPM{\hexnumber\upmath@group}
      \new@mathgroup\amsa@group
      \define@mathgroup\mv@normal\amsa@group{msa}{m}{n}
      \define@mathgroup\mv@bold\amsa@group{msa}{m}{n}
      \edef\AMSa{\hexnumber\amsa@group}
      \makeatother
      \mathchardef\upi="0\UPM19
      \mathchardef\umu="0\UPM16
      \mathchardef\upartial="0\UPM40
      \mathchardef\leqslant="3\AMSa36
      \mathchardef\geqslant="3\AMSa3E

       \let\le=\leqslant

    \fi
  \fi
\fi 

\ifnfsstwo
  \DeclareMathAlphabet{\mathbfit}{OT1}{cmr}{bx}{it}
  \SetMathAlphabet\mathbfit{bold}{OT1}{cmr}{bx}{it}
  \DeclareMathAlphabet{\mathbfss}{OT1}{cmss}{bx}{n}
  \SetMathAlphabet\mathbfss{bold}{OT1}{cmss}{bx}{n}
  \ifAMStwofonts
    \ifCUPmtlplainloaded \else
      \DeclareSymbolFont{UPM}{U}{eur}{m}{n}
      \SetSymbolFont{UPM}{bold}{U}{eur}{b}{n}
      \DeclareSymbolFont{AMSa}{U}{msa}{m}{n}
      \DeclareMathSymbol{\upi}{0}{UPM}{"19}
      \DeclareMathSymbol{\umu}{0}{UPM}{"16}
      \DeclareMathSymbol{\upartial}{0}{UPM}{"40}
      \DeclareMathSymbol{\leqslant}{3}{AMSa}{"36}
      \DeclareMathSymbol{\geqslant}{3}{AMSa}{"3E}

       \let\le=\leqslant

    \fi
  \fi
\fi 

\ifCUPmtlplainloaded \else
  \ifAMStwofonts \else 
    \def\upi{\pi}
    \def\umu{\mu}
    \def\upartial{\partial}
  \fi
\fi
\def\etal{{\it et al. }}

\begin{document}

\title[
NGC 3597 Globular Clusters
]
{
Imaging of the Merging Galaxy NGC 3597 and its 
Population of Proto--Globular Clusters
}

\author[
Duncan A. Forbes and G. K. T. Hau
]
{
Duncan A. Forbes$^1$ and G. K. T Hau$^{2}$\\
$^1$School of Physics and Astronomy, 
University of Birmingham, Edgbaston, Birmingham B15 2TT \\
(E-mail: forbes@star.sr.bham.ac.uk)\\
$^2$ Departamento de Astronom\'\i a y Astrof\'\i sica, P.~Universidad
Cat\'olica, Casilla 104, Santiago 22, Chile\\
(E-mail: ghau@astro.puc.cl)\\
}

\pagerange{\pageref{firstpage}--\pageref{lastpage}}
\def\LaTeX{L\kern-.36em\raise.3ex\hbox{a}\kern-.15em
    T\kern-.1667em\lower.7ex\hbox{E}\kern-.125emX}

\newtheorem{theorem}{Theorem}[section]

\label{firstpage}

\maketitle

\begin{abstract}

\noindent
We present wide field-of-view near--infrared imaging from the NTT and
very deep optical imaging from the HST of the young merging galaxy NGC
3597. The morphology of the galaxy and the properties of the
newly formed proto--globular clusters (PGCs) are examined.  
Our K band data reveals the presence of a
second nucleus, which provides further evidence that NGC 3597 is the result
of a recent merger. Combining new K band photometry with optical
photometry, we are able for the first time to derive a unique age for the
newly formed PGCs of a few Myrs. 
This is consistent with the galaxy starburst age of $\le$ 10
Myrs. From deep HST imaging, we are able to probe the luminosity function
$\sim$ 8
magnitudes fainter than normal, old globular clusters, and confirm that the
PGCs have a power--law distribution with a slope of $\sim$ --2. 

\end{abstract}


\section{Introduction}

In the 1970's, the rich globular cluster (GC) populations seen around
elliptical galaxies were used to argue against the idea that
ellipticals formed from the simple merger of two spirals (van den
Bergh 1975). The suggested solution to this problem was the creation
of new GCs from the gas associated with the progenitor galaxies
(Schweizer 1987; Ashman \& Zepf 1992). 
The {\it Hubble Space Telescope} (HST), with its high spatial
resolution, has indeed detected proto--globular cluster candidates in
several merging galaxies. Over a dozen such systems have now been
observed with HST (e.g.  Holtzman \etal 1992; Whitmore \etal 1993;
Whitmore \& Schweizer 1995; Schweizer \etal 1996; Holtzman \etal 1996;
Miller \etal 1997; Zepf \etal 1999; Carlson \etal 1999).  
These studies suggest that all mergers involving
gaseous systems 
create proto--globular clusters (PGCs).
However there are several outstanding issues concerning these PGCs, 
such as the total number created, their luminosity (mass) function, 
their destruction rate and the overall specific frequency of the final
system (e.g. Kissler--Patig, Forbes \& Minniti 1998; Brodie \etal 1998; 
Zepf \etal 1999). 

Perhaps the first detailed study of PGCs in a
merging system was that of Lutz (1991). Lutz presented optical imaging
and spectroscopy of NGC 3597 (AM 1112-232). 
Classified as an S0pec, it reveals an
extended structure with plumes but the main body of the galaxy 
resembles an r$^{1/4}$--like surface brightness profile. 
Using the ESO 2.2m telescope, under $\sim$1.3$^{''}$ seeing
conditions, Lutz detected and measured photometry for 31 unresolved
objects around the galaxy. About half of these were suggested to be
young, blue PGCs.  Following this ground--breaking
work by Lutz, Holtzman \etal (1996) re--observed NGC 3597 using the
WFPC2 camera on HST. They detected over 70 PGC candidates. The bulk of
these had similar colours suggesting a single age and metallicity
population. However from their V--R colours alone they were unable to
identify a unique age for the PGCs. 

Here we present K band (2.2$\mu$m) imaging of NGC 3597 taken under photometric
conditions with $\sim$0.6$^{''}$ seeing.  The near--infrared has the advantage
of being much less affected by dust extinction (A$_K$ $\sim$ 0.1
A$_V$). We also incorporate deep B and R band data from the HST
archive. The
near--infrared photometry, 
when combined with optical magnitudes, provides a powerful
constraint on the age of PGCs. After completing our study, Carlson \etal
(1999) published a study of NGC 3597 using the same deep HST
imaging. However as described below, the deep B and R data alone do not
provide a significant improvement over the Holtzman \etal V and R band data
in determining the age of the PGCs. Our main contribution is the addition
of new K band photometry which allows the age of the PGCs to be uniquely
constrained for the first time. The K band images also reveal a second
nucleus, not obvious in the shorter wavelength HST images, but apparent
from radio imaging. 


\section{Near--infrared Data}

\subsection{Observations and Data Reduction}

K short (Ks; $\lambda / \Delta \lambda$ = 2.162/0.275) 
images of NGC 3597 were taken with the infrared camera SOFI on
the ESO NTT over 2 nights in 1999 March.  The data were obtained 
under photometric conditions and excellent 
seeing (0.6$^{''}$). SOFI has a field-of-view of $\sim$ 
5 $\times$ 5 $^{'}$ and a pixel size of 0.292$^{''}$. 
A total of 220 $\times$ 30sec exposures were
taken to give a total effective exposure time of 110min. 
Between each 30sec exposure, the telescope position was 
offset by $\sim$1$^{'}$. Standard stars from the list of
Persson \etal (1999) were also taken before and after each 
block of galaxy observations. 

Flat--fields were made by median combining the standard star images. These 
were determined to be superior to the dome and twilight flats. 
After subracting dark currents and flat--fielding, the 
individual images were carefully shifted and median 
combined to form a single final Ks image of NGC 3597. 
The sky background of the final image is flat to $\sim$2\%. 

From our standard star images taken directly before and after the galaxy, 
we have determined a Ks zero point of 
22.23 $\pm$ 0.03, where the uncertainty represents the errror on the mean from 
several different images.  
We will not attempt to 
transform our Ks magnitudes to standard K band values, but 
simply note that they are similar and have been defined so that Ks = K for 
A0 standard stars (Persson \etal 1999). Hereafter we simply use K to refer
to our K short band data. 

\subsection{Galaxy Modelling}

We have modelled the galaxy using the STSDAS task {\it ellipse.} The central
position and the position angle of the model were fixed, but the
ellipticity was allowed to vary. Stars and bright PGCs were masked out. 
We also used sigma clipping to exclude
the most deviant data values from the model fit. 
Subtraction of the resulting galaxy model showed
that it was a reasonable representation except in the very inner regions. 
The K surface brightness profile for the model is shown in Fig. 1. The
outer regions are close to an r$^{1/4}$ surface brightness profile, as
found in most elliptical galaxies. 
The total magnitude of the galaxy, from a curve-of-growth analysis, is K =
11.75 $\pm$ 0.05. We later use the model--subtracted image for the detection of
globular clusters.

\section{Results and Discussion}

\subsection{The Host Galaxy}

The basic properties of NGC 3597 are summarised in Table 1. Coordinates of the
galaxy centre are taken from an HST R band (F702W) image, but are probably
only accurate to within $\pm$1$^{''}$ due to dust. However this 
position is
consistent with the position of the brightest radio source (van Driel
\etal 1991). The distance of 49 Mpc is based on a Virgocentric inflow
corrected velocity of 3513 km s$^{-1}$ and 
H$_o$ = 75 km s$^{-1}$ Mpc$^{-1}$. 
At this distance 1$^{''}$
corresponds to 240 pc. The galaxy is fairly luminous at all wavelengths, 
and is
undergoing a vigourous starburst (e.g. Lutz 1991; Kim \etal 1995;
Rephaeli \etal 1995). 

In Fig. 2 we show the inner regions of our K image of NGC 3597. 
We also identify the location of the two radio sources mapped at 6cm
by van Driel \etal (1991), which correspond spatially with the bright 
central source and a second source in our K image. 
The galaxy inner regions 
reveal that the bright central source is connected via a 
`bridge' to the second source about 3.8$^{''}$ (0.9 kpc) 
to the West. This could be simply a
super starcluster, but it is not particuarly bright at optical
wavelengths. At radio wavelengths, it appears as a connected second source
with a similar flux (S$_{6cm}$ = 18.0 mJy compared to the central source of
29.1 mJy). Thus it appears quite likely that the central source and the 
second source, seen in the
K band and at 6cm, are the nuclei of the merging galaxies. The dynamical
timescale for these nuclei to merge is less than 10$^6$ yrs, so we appear
to be witnessing NGC 3597 in the very final stages of nuclear coascelence. 

With a projected 
nuclear separation of 0.9 kpc, NGC 3597 is at a similar evolutionary
stage to NGC
3256 (Norris \& Forbes 1995) and NGC 6240 (van der Werf \etal 1993) with
separations of 0.73 and 0.90 kpc respectively. The ratio of far--infrared
to H$_2$ mass is a measure of star formation efficiency, and has been shown
to increase as the nuclear separation of two merging galaxies reduces to
zero (Gao \& Solomon 1998; Georgakakis, Forbes \& Norris 1999). 
NGC 3597 has a L$_{FIR}/M_{H_2}$ of 25.6
(Wiklind \etal 1995), comparable to 20 for NGC 3256 and 30 for NGC 6240. 
This further supports the case that NGC 3597 is near the nuclear
coascelence stage of a merger sequence. 

Beyond
our faintest isophote, deep optical images reveal plumes 
($\mu _R \sim 25$ mag arcsec$^{-2}$)
to the NW and SW (Lutz 1991). 
As NGC 3597 is relatively isolated,  
the outer morphological disturbance is almost certaintly due to a merger of
some sort. Indeed the morphology of the plumes resembles the merger
simulation shown in Barnes (1998) at 300 Myr after pericentre, and 
just before nuclear merger.

The galaxy is currently undergoing a starburst. Optical spectra
(e.g. Lutz 1991; Kim \etal 1995) reveal emission lines indicative of
HII regions. However, as well as the current star formation activity,
there is also evidence that the burst is not instantaneous but has
proceeded for some time.  Optical spectra clearly show weak Hydrogen
absorption lines (along with the Hydrogen emission) indicative of an
earlier phase of star formation.  The radio spectral index is 
$\alpha$ = --0.84 indicating that the emission is not dominated
by HII regions but rather is due to 
non--thermal synchrotron emission from SNRs
(Smith \& Kassim 1993).  Both the current and post starburst contribute to
the far--infrared luminosity of L$_{FIR}$ $\sim$ 7 $\times$ 10$^{10}$
L$_{\odot}$ which is consistent with
the well known far--infrared vs radio correlation. 
Kim \etal (1995) estimated
an H$\beta$ absorption EW of 2.5~\AA ~ from their optical spectrum. Such an
EW occurs in the first 10 Myrs of a starburst (e.g. Bruzual \& Charlot
1993). This indicates that the starburst is very recent, starting 
less than 10 Myr ago and continuing to the present day. 
We compare this age estimate to that of the 
PGCs in the next section. 

Wikland \etal (1995) did not detect HI gas in NGC 3597 but only placed a
relatively high upper limit of M$_{HI} < 2.6 \times 10^{10}$
M$_{\odot}$. Based on the FIR luminosity and a Galactic gas-to-dust ratio,
Lutz (1991) estimated an HI gas mass of 3 $\times$ 10$^{8}$ M$_{\odot}$,
i.e. almost a factor of one hundred below the current Wikland \etal limit. 
It would be interesting to place tighter limits on the HI mass
and investigate the possibility that atomic hydrogen gas is being
compressed during the merger into molecular gas.  

Mihos \& Hernquist (1994) have modelled the star formation history of equal
mass spiral galaxies. They found that for systems without strong bulges 
(e.g. Sc spirals), most of the merger--induced star formation occurs at the
pericentre encounter over a timescale of a few Myrs. 
However the presence of a bulge inhibits gas flow and
delays the main starburst until nuclear coalescence. 
This starburst also lasts only a
few Myrs. The age of the starburst in NGC 3597 (i.e. $\le$ 10 Myrs) and the
galaxy morphology (i.e. two nuclei and a common r$^{1/4}$ like envelope)
suggests that the latter situation is the correct one. Thus it appears
that NGC 3597 is likely the product of two near equal mass spirals, at
least one of which contained a strong bulge. Furthermore we are currently 
witnessing
it at the nuclear merger stage, which is also the period of its most
intense star
formation activity.

\subsection{The Proto--Globular Clusters} 

Our K image of NGC 3597 reveals a number of bright unresolved
sources. The vast bulk of these are PGCs. 
They were first noticed by Lutz (1991) and later
re--observed with HST (Holtzman \etal 1996; Carlson \etal 1999). 
In the discussion that
follows we will assume a 
Galactic extinction towards NGC 3597 of A$_V$ = 0.12 (A$_B$ = 0.16, A$_R$ =
0.08, and A$_K$ = 0.01) as used by Holtzman \etal ~Although the galaxy
shows a dust lane to the North of the nucleus, the internal extinction
towards most of the PGCs is probably not large given the relatively uniform
V--R colors noted by Holtzman \etal (see their figure 6).

Lutz (1991) tabulated 31 sources. From Holtzman \etal (1996), it appears
that 9 of them are {\it bona fide} PGCs, i.e. numbers 10, 11, 13, 14, 17,
19, 21, 23 and 24 from his table (the other 22 sources are either
foreground stars or background galaxies). 
The mean B--V colour for these 9 sources, 
corrected for
Galactic extinction, is 0.40 with a 1$\sigma$ error on the mean of 0.03. 
This colour and range for the PGCs is shown in Fig. 3, along with the
evolutionary track for a solar metallicity, Salpeter IMF 
single stellar population from 
Bruzual \& Charlot (1993). The IMF and metallicity of the PGCs are 
of course unknown,
so any derived ages will be a guideline only. The average PGC colour
intersects the track at about 6--10, 100 and 1000 Myrs. Thus it 
is impossible to
derive a unique age from the B--V colour alone.

Holtzman \etal (1996) used WFPC2 to obtain F555W (V) and F702W (R) 
photometry of the PGCs. They detected 72 PGCs in the PC chip 
down to V $\sim$ 25.5. 
The first source listed by Holtzman \etal (Holtz ID = 1) is
probably a star with V = 20.04 and V--K = 1.4 which suggests it is 
a G0 dwarf star. For the 60 globular clusters with V $<$ 25, the mean
colour and error on the mean is V--R = 0.34 $\pm$ 0.03 (again corrected for
Galactic extinction). This is shown in Fig. 4 along with the 
Bruzual \& Charlot (1993) evolutionary track. As with B--V, the 
V--R colours do not give a single age but several possibilities,
i.e. $\sim$ 6, 12, 80, 200 Myrs. The V--R colours do however rule out the
oldest B--V age of 1 Gyr.

Using the HST PGC positions, we measured K magnitudes using {\it daophot.}
This was carried out on the galaxy subtracted image, using 
small aperture magnitudes for which we then applied an aperture correction
(based on objects away from the galaxy centre). 
We detected 31 PGCs in our image down to K $\sim$ 22.8, although with
large errors at faint magnitudes. 
Our K magnitudes and measurement errors are listed in Table
2, along with Holtzman \etal ID number and offset from the galaxy centre. 
All of these PGCs lie within $\sim$ 20$^{''}$ of the galaxy centre
(i.e. detected in the PC chip by Holtzman \etal~).

In Table 3 we list the V magnitude, V--R colours and errors 
from Holtzman \etal (1996). The errors are photometric measurement errors,
and do not include zeropoint errors (on the order of 0.05--0.1 mag.). 
These errors have been combined in quadrature to obtain the final V--K
error. The mean colour and error on the mean is V--K = 0.95 $\pm$ 0.14. 
This range of colours is shown in Fig. 5, along with the 
Bruzual \& Charlot (1993) evolutionary track. The extra `leverage' from the
K band means that the V--K colours of the PGCs only intersect the track at
one place, i.e. around 5 Myrs. This is consistent with the youngest age
suggested by the B--V and V--R colours, and rules out the older ages at the
3$\sigma$ level. Any correction for internal extinction, or if the PGCs
had supersolar metallicities, would make the age
even less than 5 Myrs. Such an age should be regarded as somewhat
qualitative since it is depedent on unknown factors (e.g. the IMF) and on
the particular stellar population model used (i.e. Bruzual \& Charlot
1993). 

We have obtained 
deep F450W (B) and F702W (R) HST images of NGC 3597 from
the CADC HST archive. The 8 images total 5200sec in B and 5000sec in R, and
cover essentially the same area as the shallower Holtzman \etal (1996)
data.  They were
average combined using the STSDAS task {\it gcombine}, which effectively
removes cosmic rays. The high throughput of the F702W filter and the
spectral energy distribution of globular clusters means that the R image
will be somewhat deeper than the B image. Nevertheless both images should
reach magnitudes typical of the brighter GCs associated with the 
original, old 
population from the progenitor galaxies. 
The previous HST imaging published
by Holtzman \etal (1996) had an F702W exposure time of 1100sec and 
was thus not 
deep enough to reach the old population and only probe the bright end of
the new population. This data was also recently used by Carlson \etal
(1999), and reaches
about 1.6 mags deeper in R than the Holtzman \etal study. 

In order to locate sources, we used {\it daofind} with a conservative 
4$\sigma$ per pixel detection criterion. For the PC chip, we fit the
galaxy isophotes using {\it ellipse} and subtracted off a model before
running {\it daofind.} Because of the confused nature in the central
regions in the B image, 
due to dust and young star formation, we have excluded the central
9$^{''}$ radius from the automatic detection. 
Photometry on all four chips was measured using {\it
phot} with a 2 pixel radius aperture. Aperture corrections to 0.5$^{''}$ and
zero points were taken from Holtzman \etal (1995, 1996). 
The objects on the PC chip are marginally resolved. Holtzman \etal (1996)
showed that the bulk of objects required a typical correction of 0.22$^m$, in
addition to the above aperture correction, to
include all of the light. Variations with the size of
the object and its position on the chip, could introduce another source of
error on the order of 0.1$^m$. However the correction to total light
affects the two filters almost equally, so there is little error introduced
in the final B--R colour. 

The resulting object lists were then checked visually on the image display to
exclude the very few remaining obvious bright stars and background galaxies. 
At this stage, our candidate GC list contained 292 sources from all four
chips (excluding the central 9$^{''}$ of the PC chip). A
colour--magnitude diagram is shown in Fig. 6. We have adopted selection
cuts of --0.5 $<$ B--R $<$ 2.0 and B $<$ 27. The colour cuts correspond to
the full range expected of GCs older than a Myr (Bruzual \& Charlot
1993). The faint magnitude limit was chosen to avoid any colour bias in our
sample. Within these selection criteria we have 239 objects, the vast bulk
of which will be {\it bona fide} GCs (as our contamination rate is $\sim$
5\%). 

A histogram of B--R colour is shown in Fig. 7. The GCs have an average colour
of B--R = 0.66 $\pm$ 0.03 (error on the mean). 
The spread in colour is consistent with photometric errors suggesting that
the PGCs are close to a single age and metallicity population. 
This mean value is shown on
the evolutionary tracks of Bruzual \& Charlot (1993) in Fig. 8. From B--R
colour alone the GCs may have several possible 
ages, but from our K band imaging
above it is clear that the correct age is $\sim$ 5 Myrs. Thus the majority
of detected objects in our deep HST images are PGCs. 

Have we detected any GCs from the progenitor galaxies ? Milky Way
GCs have a mean B band luminosity function that is roughly Gaussian with a
peak at M$_B$ $\sim$ --6.6, with the brightest GC ($\omega$ Cen) having 
M$_B$ = --10.7. These correspond to B = 26.85 and 22.75 respectively, at
the distance of NGC 3597. Uncertainty in the distance makes these
magnitudes uncertain by about $\pm$ 0.2$^m$. 
Milky Way GCs have an average extinction--corrected colour of (B--R)$_o$
$\sim$ 1.2. 
Examination of the colour--magnitude diagram reveals 
32 GCs in the expected colour and magnitude range (shaded region in Fig. 6), 
although perhaps half of these will simply be PGCs with apparently red 
B--R colours due to photometric errors. If we assume that say 20 are 
true old GCs then we can crudely estimate the total population of 
progenitor GCs. The HST images cover about half of the area out to a 
galactocentric radius of 125 arcsec, suggesting 40 GCs within this radius. 
Our HST data only reach to B $\sim$ 26.5, which
is slightly less than the expected peak at B = 26.85. This suggests that 
we are sampling about 40\% of the GC population in magnitude terms. 
Thus another correction of $\sim$2.5$\times$ is required, giving a total 
old GC population of about 100. 
This crude calculation could easily be 
in error by a factor of two, but is not vastly different to the halo 
GC population of the Milky Way (i.e. $\sim$120). Unfortunately such a 
calculation does not constrain the progenitor types. 
Deeper B band observations, over a somewhat 
wider field of view, would provide better constraints on the original GC
systems. 

Globular cluster 
luminosity functions in old ellipticals show a Gaussian log--normal
distribution, with a peak, or turnover, magnitude 
and fewer GCs at low luminosity (mass). However, for the PGC systems
studied to date 
the luminosity function does not peak but continues as a power--law
down to low luminosities. Published slopes include 
--1.78 (NGC 4038/9; Whitmore \& Schweizer 1995), --2.1 (NGC
3921; Schweizer \etal 1996), --1.84 (NGC 7252; Miller
\etal 1997) and --1.8 (Zepf \etal 1999). For the closer systems, the PGCs
in the PC chip are marginally resolved giving sizes consistent with GCs
rather than open clusters (which also have a power--law distribution). 
Recently Carlson \etal (1999) have examined the luminosity function for 
NGC 3597. They found a slope of --2.0 to be a reasonable representation. Here 
using the same HST data, we also probe the R band luminosity function. 

Before probing the GC luminosity 
we need to estimate the incompleteness of our R band detections. This
was achieved by simulating GCs using the {\it addstar} task, and measuring
the detected fraction as a function of R band magnitude. 
An actual WFC chip was used to reproduce the correct noise
characteristics. Care was taken to avoid any blending of artifical GCs. 
The resulting 
completeness function for the WFC chips 
is shown in Fig. 9. Our 50\% incompleteness level is
at R $\sim$ 26.8. In Fig. 10 we show the R band luminosity function in
log--log space, for the three WFC chips 
and corrected for incompleteness. It is very similar to that shown in figure 
10 of Carlson \etal (1999). 
A Milky Way like GC system with a peak magnitude of M$_R$ = --8, would 
have an apparent magnitude of R $\sim$ 25.5 at the distance of NGC 3597 and
be an additional 5.5 magnitudes brighter if they were 5 Myr old 
instead of 12 Gyr old (assuming a solar
metallicity population). This gives an expected peak of R $\sim$ 20, with
the faint (lower mass) limit of R $\sim$ 23. 
By constrast, the
observed PGC luminosity function 
continues to rise to faint magnitudes (R $\sim$ 28) with a power--law
like slope of around --2. At the faint end of our luminosity function,
assuming a Salpeter IMF, derived masses correspond to 
$\sim$ 200 M$_{\odot}$. 
At 8 magnitudes fainter than the expected peak,
these data probe deeper than any previously published study of
PGCs. We note that similar mass calculations by Carlson \etal (1999) assumed 
an age of $\sim$500 Myrs, which means that their masses are overestimates. 
Thus, like other merging galaxies, the low mass 
PGCs in NGC 3597 must be destroyed (e.g. by tidal disruption or
evaporation) over time if their luminosity functions are to 
resemble those of old ellipticals (see Zepf \etal 1999). 

Elmegreen \& Efremov (1997) have proposed that a universal mechanism exists
for GC formation. In their model, GCs form in high pressure regions 
with an assumed power--law mass distribution of slope $\sim$ --2. 
They suggest that a GC system is 
formed
initially without a characteristic mass, but as low mass GCs are destroyed
over time a characteristic mass develops. Over $\sim$ 10 Gyr, the
luminosity function grows to resemble that of the Milky Way GC system with a 
characteristic mass corresponding to M$_R$ $\sim$ --8.0  

In our K image there are many unresolved sources beyond the central 
$\sim20^{''}$ (i.e. the region covered by Holtzman \etal 1996). We used
{\it daofind} to find all sources 3$\sigma$ above the background noise. 
After rejecting 3 obvious bright stars, 1 galaxy and objects fainter than 
with K = 23, we were left with a list of 142 candidate PGCs. These objects
are listed in Table 4 along with their K magnitudes and photometric
errors. Some of the brighter 
objects will be foreground stars, and some will be compact
background galaxies. Indeed, after examining the area in common with the
deep HST
images, we have excluded a further three stars and two galaxies (as noted in
Table 4). As the HST images do not cover the whole area of our K image, 
we are unable
to confidently remove all stars and galaxies. 
In the absence of overlapping HST data (or optical magnitudes), it is
difficult to make conclusive statements about these outer objects, but many
will be PGCs.

\section{Concluding Remarks}

The combination of near--IR and optical imaging from the NTT and HST
telescopes has provided new insights into the merging system NGC 3597. For
the host galaxy we confirm an r$^{1/4}$ like surface brightness profile in
the outer regions and discover the presence of two closely separated
($\Delta$r $\sim$ 0.9 kpc) nuclei. The two nuclei seen in our K band image
correspond to those seen at radio wavelengths and are probably the nuclei
of the progenitor galaxies at the final stages of coalescence. Various
properties of the galaxy starburst suggest that the burst occured less than
10 Myrs ago and may have involved at least one early type spiral. 

A number of proto--globular clusters (PGCs) are identified in NGC 3597 with 
average colours of V--K = 0.95. The extra leverage provided by our K band
photometry allows us to derive a unique age for the PGCs of a few Myr. 
From deep B and R HST imaging we detect $\sim$ 300 globular clusters, 
the vast bulk of which are PGCs. We derive an R band luminosity function
which reaches 8 magnitudes fainter than the expected characteristic mass of
a standard old globular cluster luminosity function, and confirm 
the slope of the
luminosity function to be about --2.
A small number of globular clusters have properties consistent with 
Galactic ones, and we speculate that they were 
associated with the progenitor galaxies. \\

\noindent{\bf Acknowledgments}\\
We thank Richard Brown, Paul Goudfrooij 
for help and useful discussions. 
Based on observations with the NASA/ESA {\it Hubble Space Telescope} under
NASA contract NAS5-26555 and the {\it European Southern Observatory}. 
GKTH acknowledges financial support by FONDECYT under grant 1990442.\\

\vskip 1cm

\noindent{\bf References}

\noindent
Ashman, K.M., \& Zepf S.E. 1992, ApJ 384, 50 \\
Barnes, J., 1998, astro-ph/9811091\\
Brodie, J.P., Schroder L.L., Huchra J.P., Phillips A.C., Kissler-Patig M.,
   Forbes D.A., 1998, AJ, 116, 691\\
Bruzual, G., \& Charlot S. 1993, ApJ 405, 538 \\
Carlson, M.N., \etal 1998, AJ, 115, 1778\\
Elmegreen B.G., Efremov Y.N. 1997, ApJ 480, 235\\
Forbes, D. A. 1998, Galaxy Interactions at Low and High
Redshift, ed. D. Sanders, in press\\
Gao, Y., \& Solomon, P. M. 1998, astro-ph/9812320\\
Georgakakis, A., Forbes, D. A., \& Norris, R. P. 1999, in preparation\\
Holtzman, J.A., et al., 1992, AJ 103, 691 \\
Holtzman, J.A., et al., 1995, PASP, 107, 156\\
Holtzman, J.A., et al., 1996, AJ 112, 416 \\
Kim, D.C., Sanders, D.B., Veilleux, S., Mazzarella, J.M., Soifer, B.T.,
1995, ApJS, 98, 12\\
Kissler-Patig, M., Forbes, D.A., Minniti, D., 1998, MNRAS, 298, 1123\\
Lutz, D.,  1991, A\&A 245, 31 \\
Mihos, C., Hernquist, L., 1994, ApJ, 390, L53\\
Miller, B.W., Whitmore, B.C., Schweizer, F., Fall, S.M., 1997, AJ, 114, 2381\\
Norris, R.P., Forbes, D.A., 1995, ApJ, 446, 594\\
Persson, S. E. Murphy, D. C., Krzeminski, W., Roth, M., \& Rieke, M. J. 
1999, AJ, 116, 2475\\
Rephaeli, Y., Gruber, D., \& Persic, M. 1995, A\&A, 300, 91\\
Schweizer, F., 1987, in Nearly Normal Galaxies, ed S. Faber (Springer, New
York), 18\\
Schweizer, F., Miller B.W., Whitmore B.C., Fall S.M., 1996 AJ 112, 1839 \\
Smith, E. P., \& Kassim, N. E. 1993, AJ, 105, 46\\
van Driel, W., van den Broek, AC., \& de Jong, T. 1991, A\&AS, 90, 55\\
Van den Bergh S., 1975, ARAA, 13, 217\\
Van der Werf, P., Genzel, R., Krabbe, A., Blietz, M., Lutz, D., Drapatz,
S., Ward, M.J., Forbes, D.A., 1993, ApJ, 405, 522\\
Whitmore, B.C., Schweizer F., Leitherer C., Borne K., Robert C., 1993,
AJ, 106, 1354 \\
Whitmore, B.C., Schweizer F., 1995, AJ 109, 960 \\
Whitmore, B.C., Miller B.W., Schweizer F., Fall M., 1997, 114, 797 \\
Wikland, T., Combes, F., \& Henkel, C. 1995, A \& A, 297, 643\\
Zepf, S.E., \etal 1999, astro-ph/9904247\\

\newpage

\begin{figure*}[p] 
\centerline{\psfig{figure=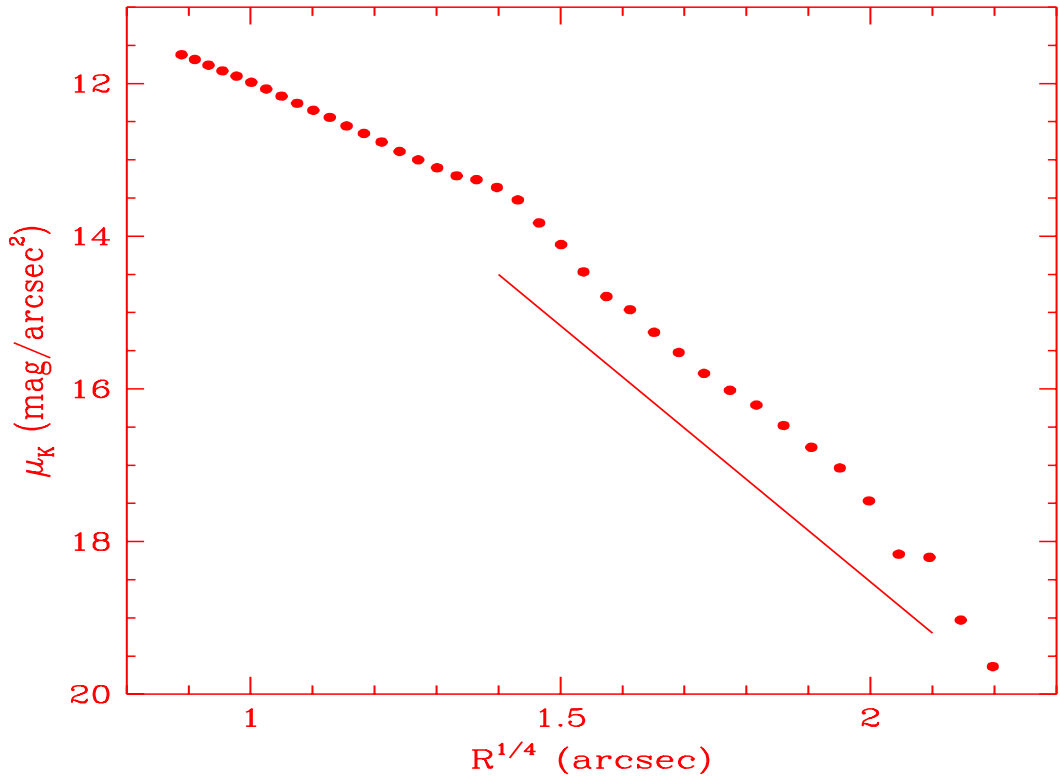,width=6in,height=6in}}
\caption{\label{fig1} K band surface brightness profile for NGC 3597. 
Errors are on the order of the symbol size. The
surface brightness in the outer parts is close to an r$^{1/4}$ profile,
i.e. a straight line. 
}
\end{figure*}

\begin{figure*}[p]
\centerline{\psfig{figure=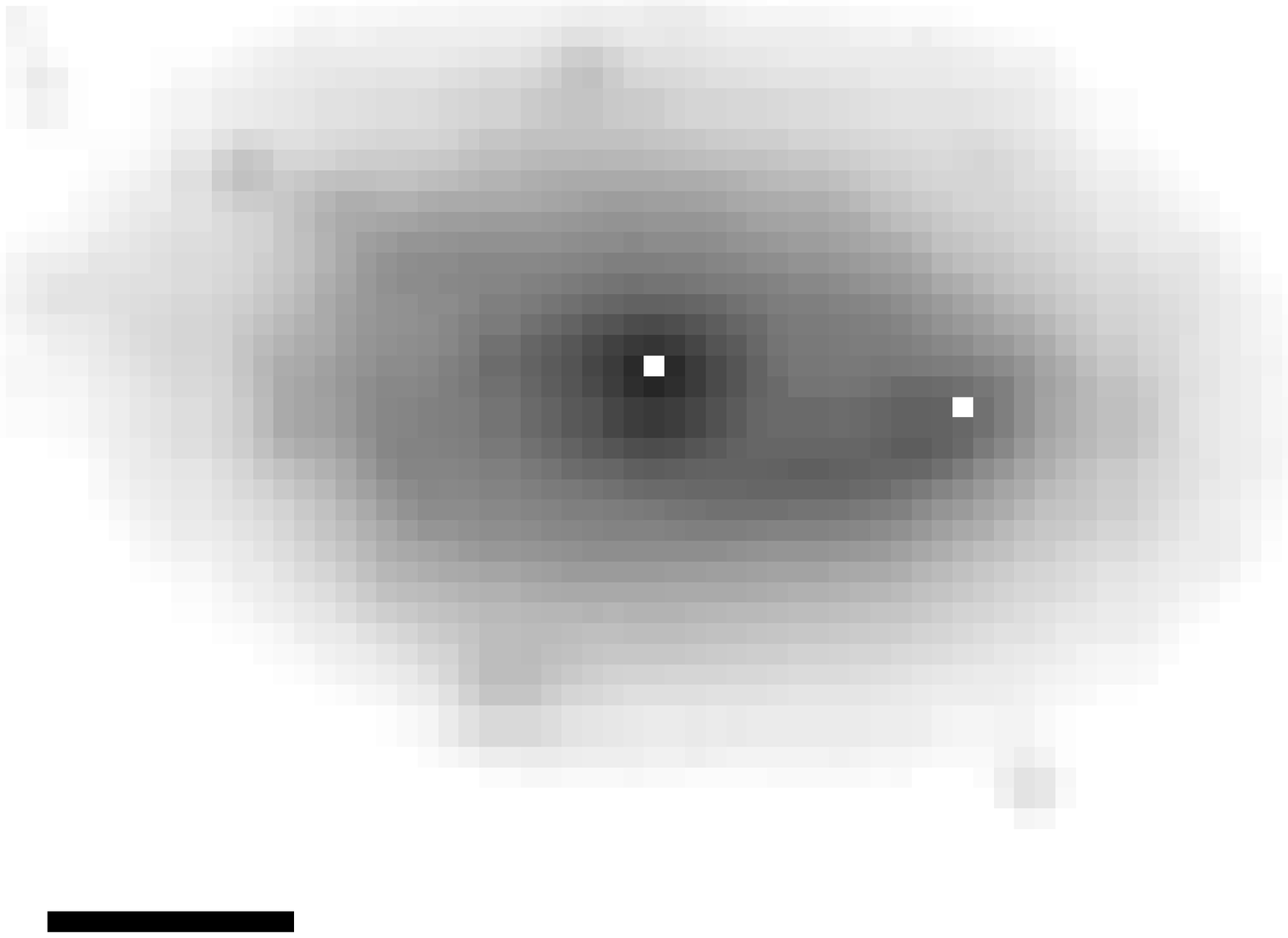}}
\caption{\label{fig2} Grey scale image of the central regions of NGC 3597
in the K band. The location of the two 6cm radio sources (which may
represent two galaxy nuclei) are indicated. The horizontal bar represents
3.5$^{''}$. North is up and East is left. 
}
\end{figure*}

\begin{figure*}[p] 
\centerline{\psfig{figure=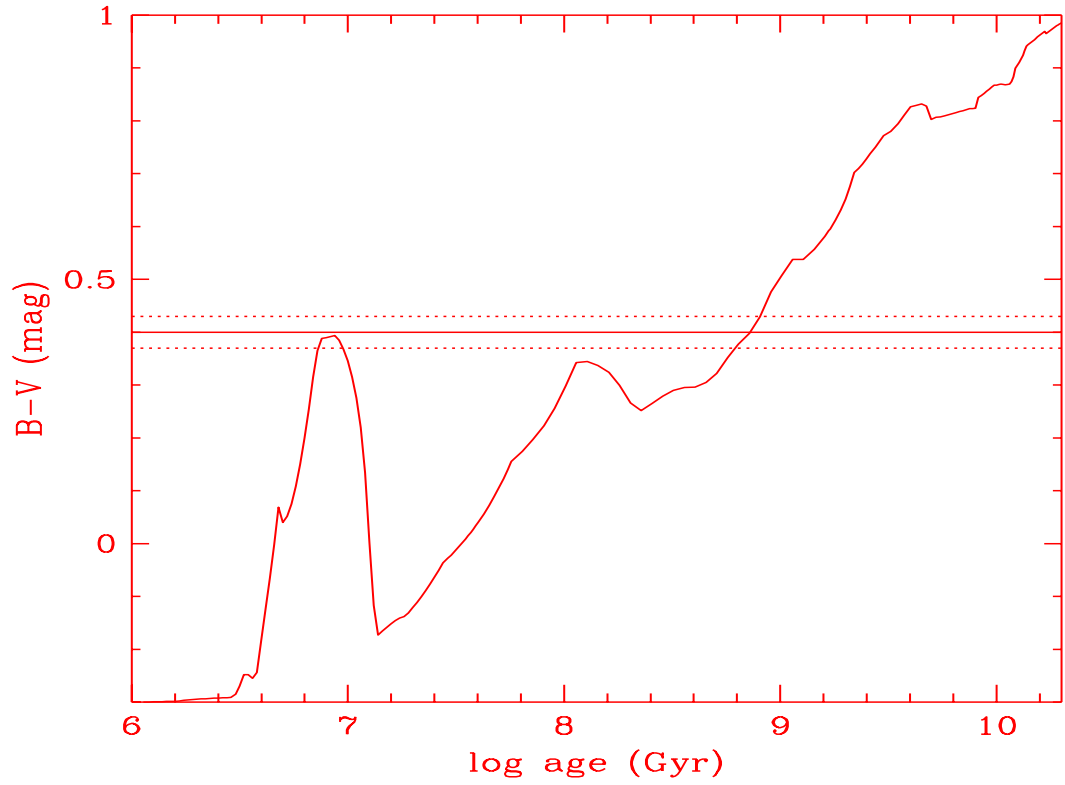,width=6in,height=6in}}
\caption{\label{fig3} B--V colour evolution for a single stellar population
of solar metallicity from Bruzual \& Charlot (1993). The horizontal line
and dashed lines show the mean B--V colour and 1$\sigma$ range for 9 
proto--globular clusters (PGCs) from Lutz (1991). The PGCs could have several
possible ages from $\sim$ 6 Myrs to 1 Gyr. 
}
\end{figure*}

\begin{figure*}[p] 
\centerline{\psfig{figure=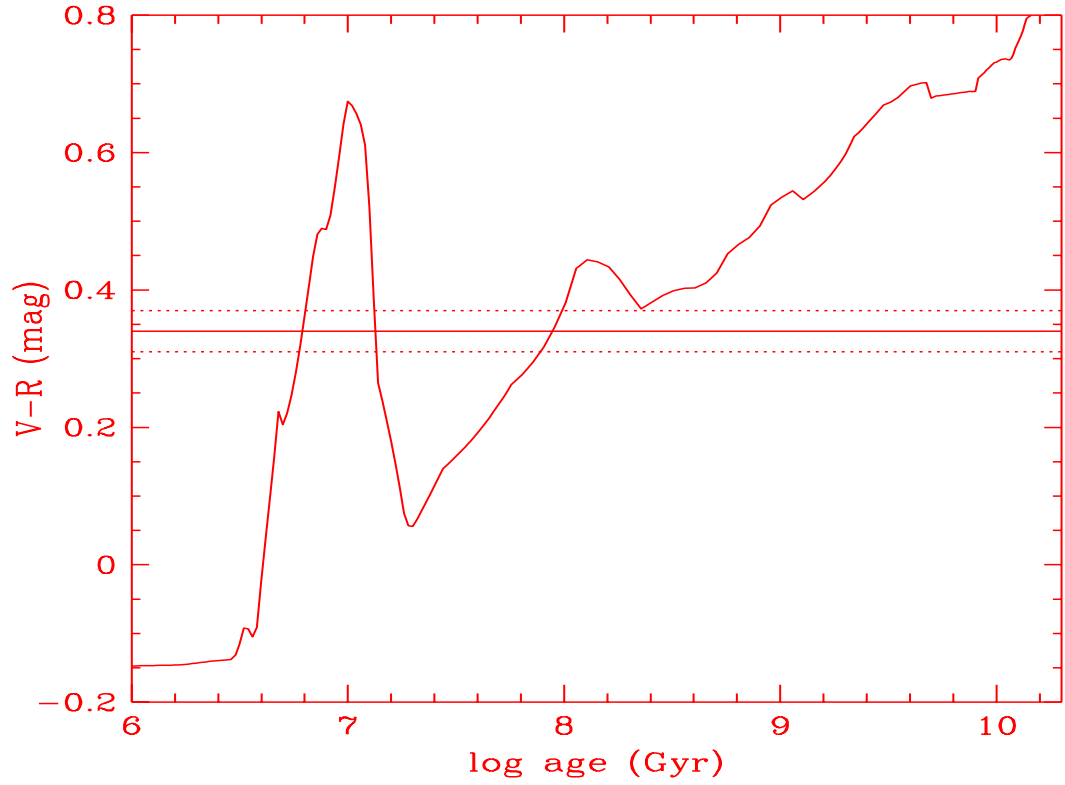,width=6in,height=6in}}
\caption{\label{fig4} V--R colour evolution for a single stellar population
of solar metallicity from Bruzual \& Charlot (1993). The horizontal line
and dashed lines show the mean V--R colour and 1$\sigma$ range for 71 
proto--globular clusters (PGCs) from Holtzman {\it et al.}  
(1996). The PGCs could have several 
possible ages from $\sim$ 6 Myrs to 200 Myr. 
}
\end{figure*}

\begin{figure*}[p] 
\centerline{\psfig{figure=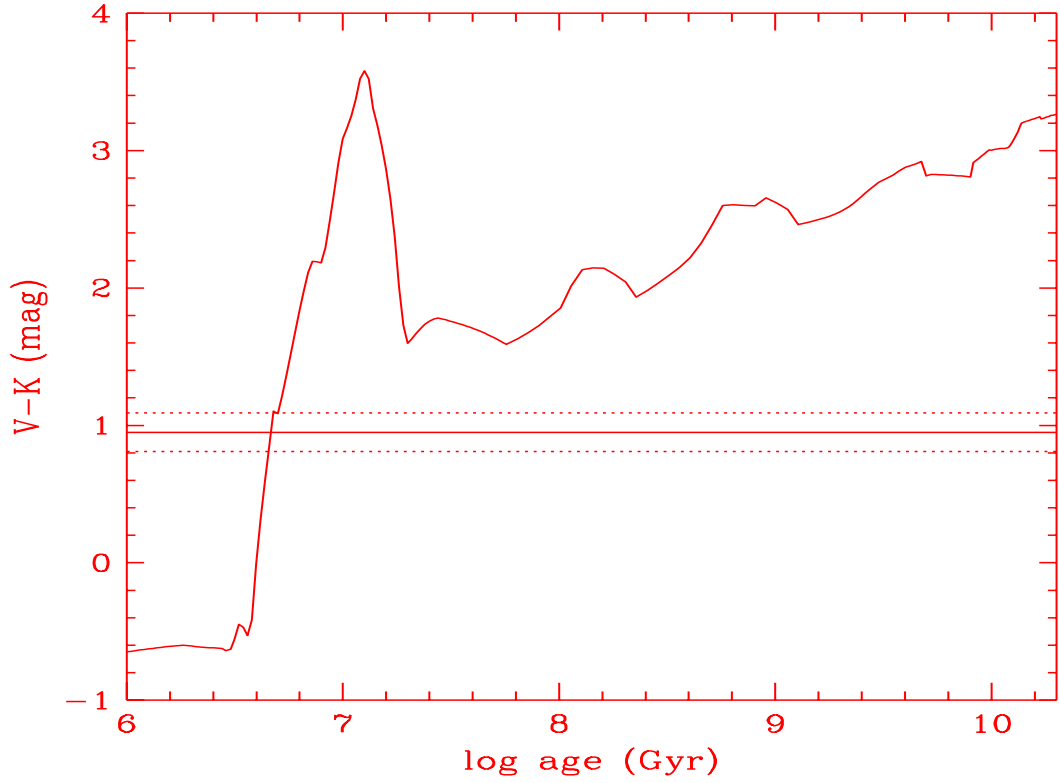,width=6in,height=6in}}
\caption{\label{fig5} V--K colour evolution for a single stellar population
of solar metallicity from Bruzual \& Charlot (1993). The horizontal line
and dashed lines show the mean V--K colour and 1$\sigma$ range for 31 
proto--globular 
clusters (PGCs) from this paper. The PGCs have an age of $\sim$ 5
Myr. 
}
\end{figure*}


\begin{figure*}[p] 
\centerline{\psfig{figure=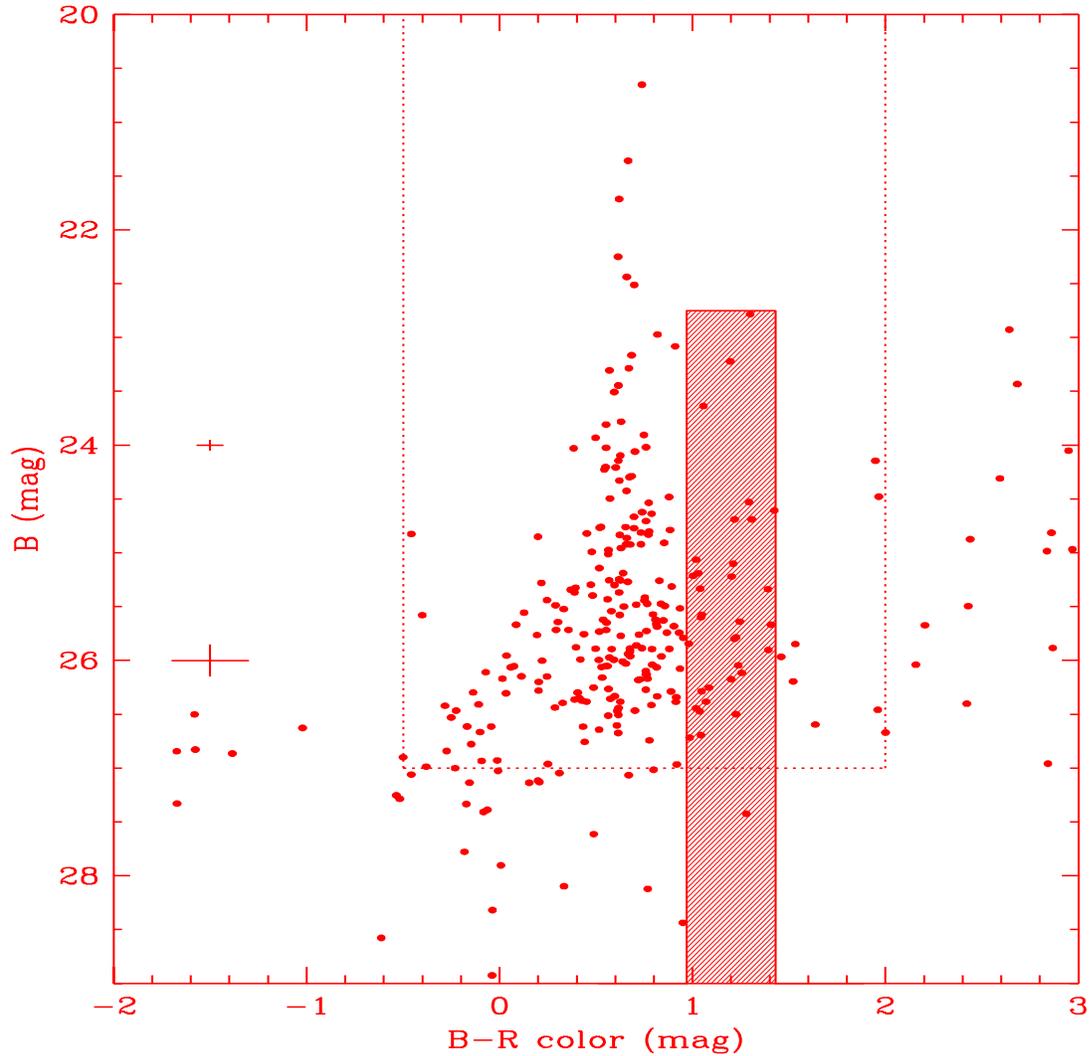,width=6in,height=6in}}
\caption{\label{fig6} Colour--magnitude diagram for the detected objects
from deep WFPC2 data. The dotted line shows the region B $<$ 27, --0.5 $<$
B--R $<$ 2.0 used to select globular clusters. Typical photometric error
bars are shown on the left. The shaded region represents the apparent 
magnitude and
colour range for Milky Way globular clusters placed at the distance of
NGC 3597. The data show that the majority of globular clusters have B--R
$\sim$ 0.66, which are identified as proto--globular clusters. There are
a few old, Milky Way like globular clusters.
}
\end{figure*}

\begin{figure*}[p] 
\centerline{\psfig{figure=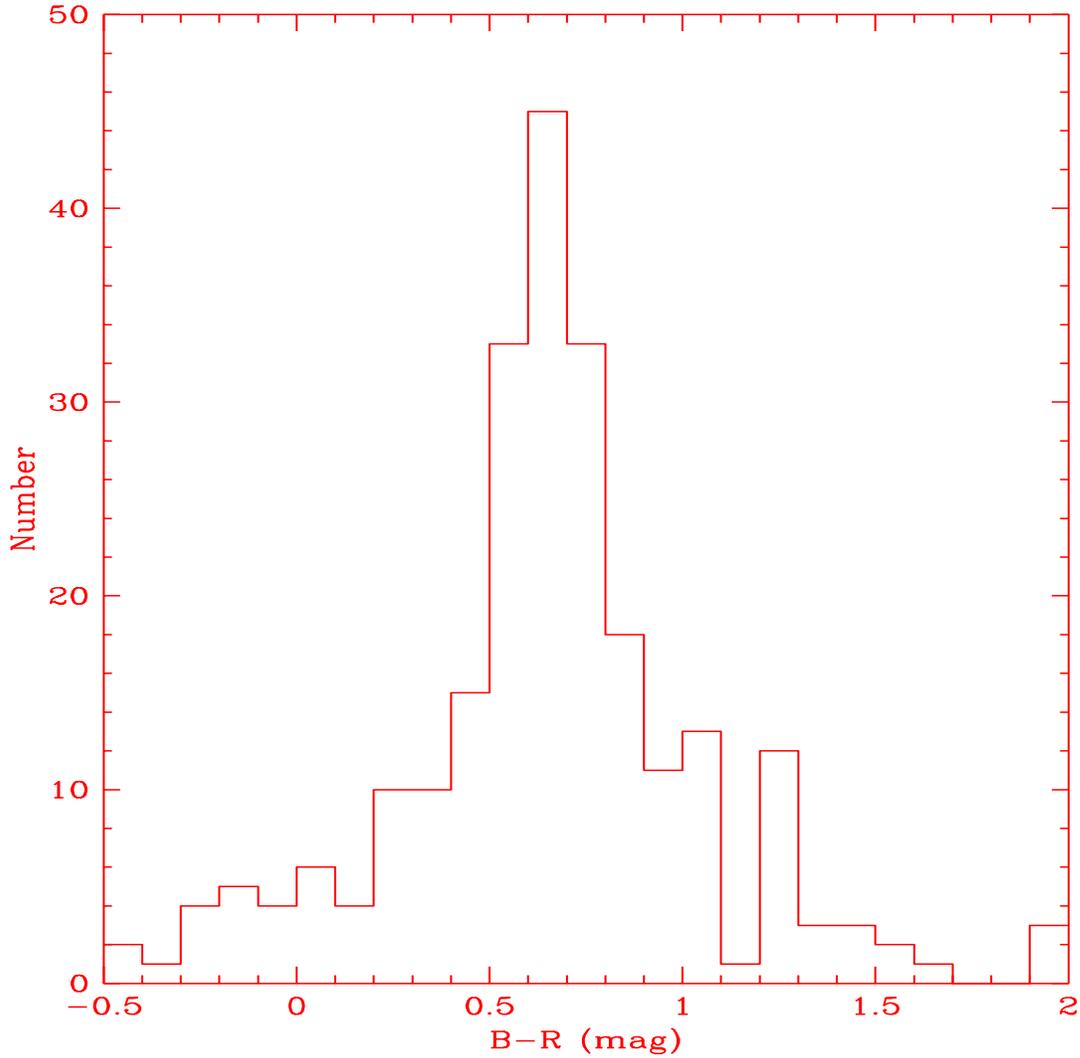,width=6in,height=6in}}
\caption{\label{fig7} B--R colour histogram of the 239 selected globular
clusters from deep WFPC2 data. The majority of globular clusters have B--R
$\sim$ 0.66, which are identified as proto--globular clusters.
}
\end{figure*}

\begin{figure*}[p] 
\centerline{\psfig{figure=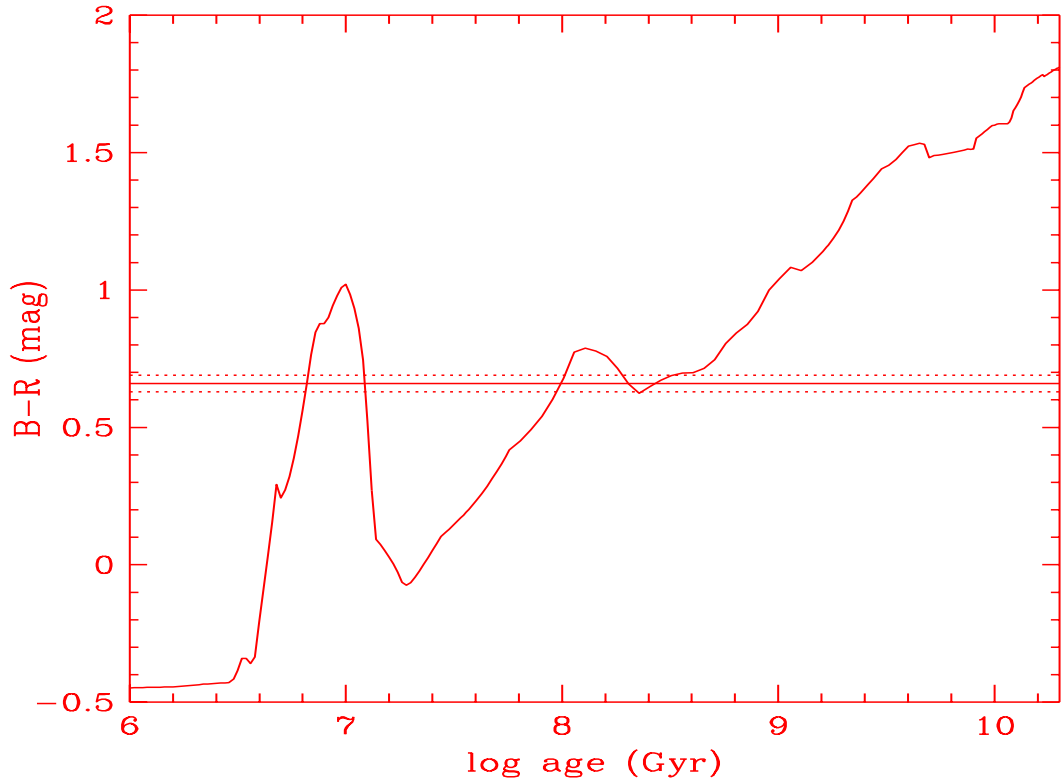,width=6in,height=6in}}
\caption{\label{fig8} B--R colour evolution for a single stellar population
of solar metallicity from Bruzual \& Charlot (1993). 
The horizontal line
and dashed lines show the mean B--R colour and 1$\sigma$ range for $\sim$239 
proto--globular clusters (PGCs) from this paper. The age of the 
PGCs is consistent with $\sim$ 5 Myr determined from V--K colours.
}
\end{figure*}

\begin{figure*}[p] 
\centerline{\psfig{figure=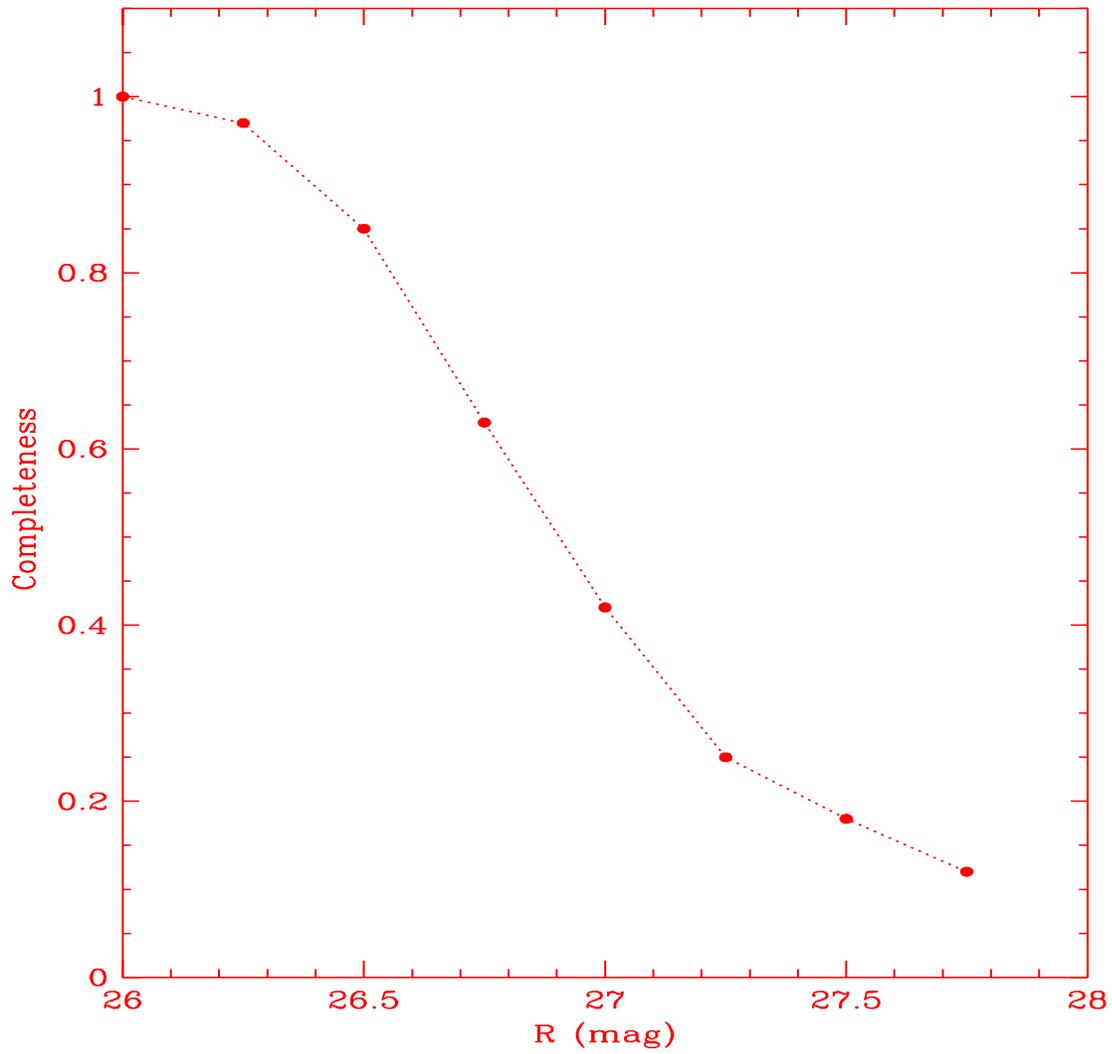,width=6in,height=6in}}
\caption{\label{fig9} Completeness function for globular clusters from
simulations. Brighter than R = 26, essentially all globular clusters are
detected. 
}
\end{figure*}

\begin{figure*}[p] 
\centerline{\psfig{figure=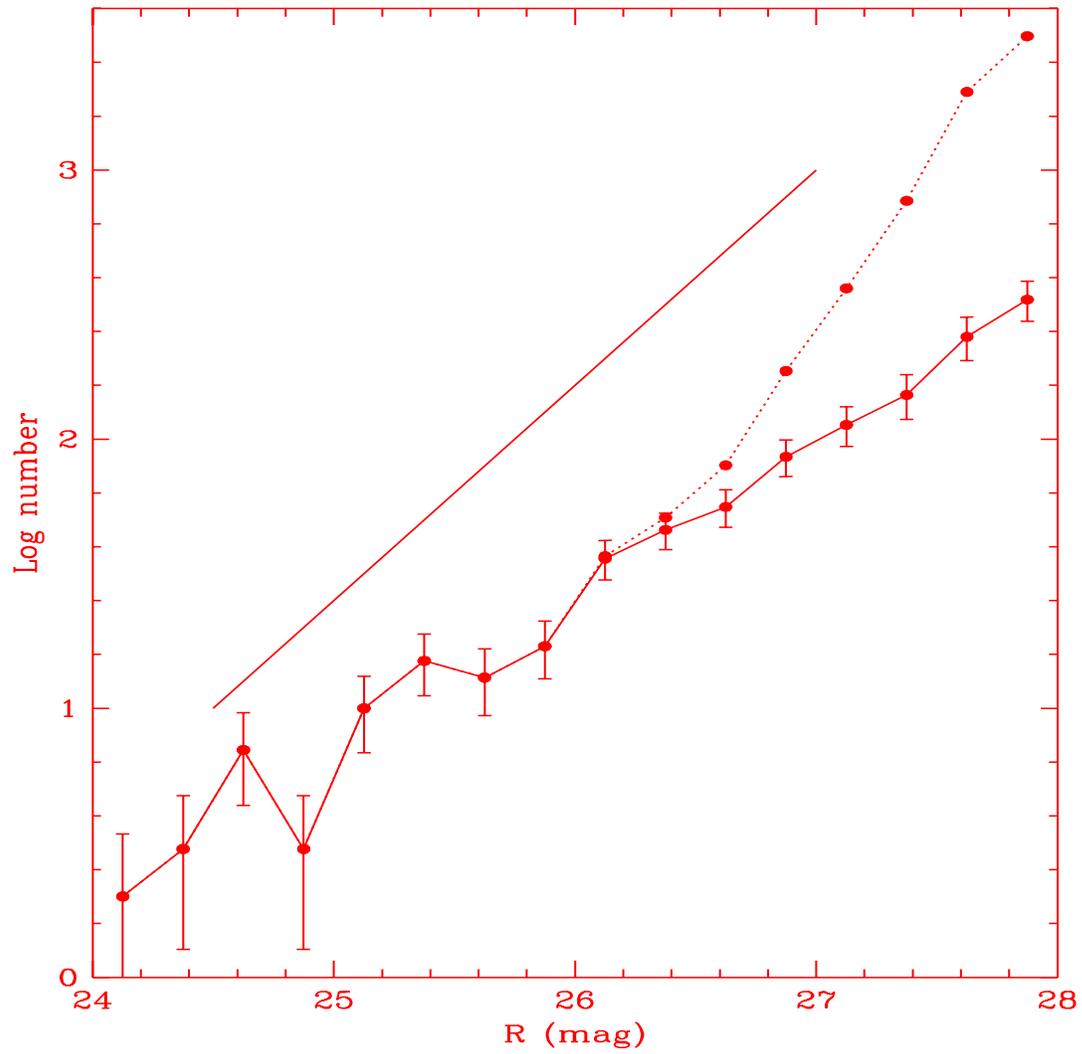,width=6in,height=6in}}
\caption{\label{fig10} Logarithmic globular cluster luminosity
function in the R band. 
The solid line shows the raw data, and the dashed line after 
correction for incompleteness. The objects brighter than R = 24 are not
shown. The straight line shows a power--law slope
of --2. 
}
\end{figure*}

\begin{figure*}[p] 
\centerline{\psfig{figure=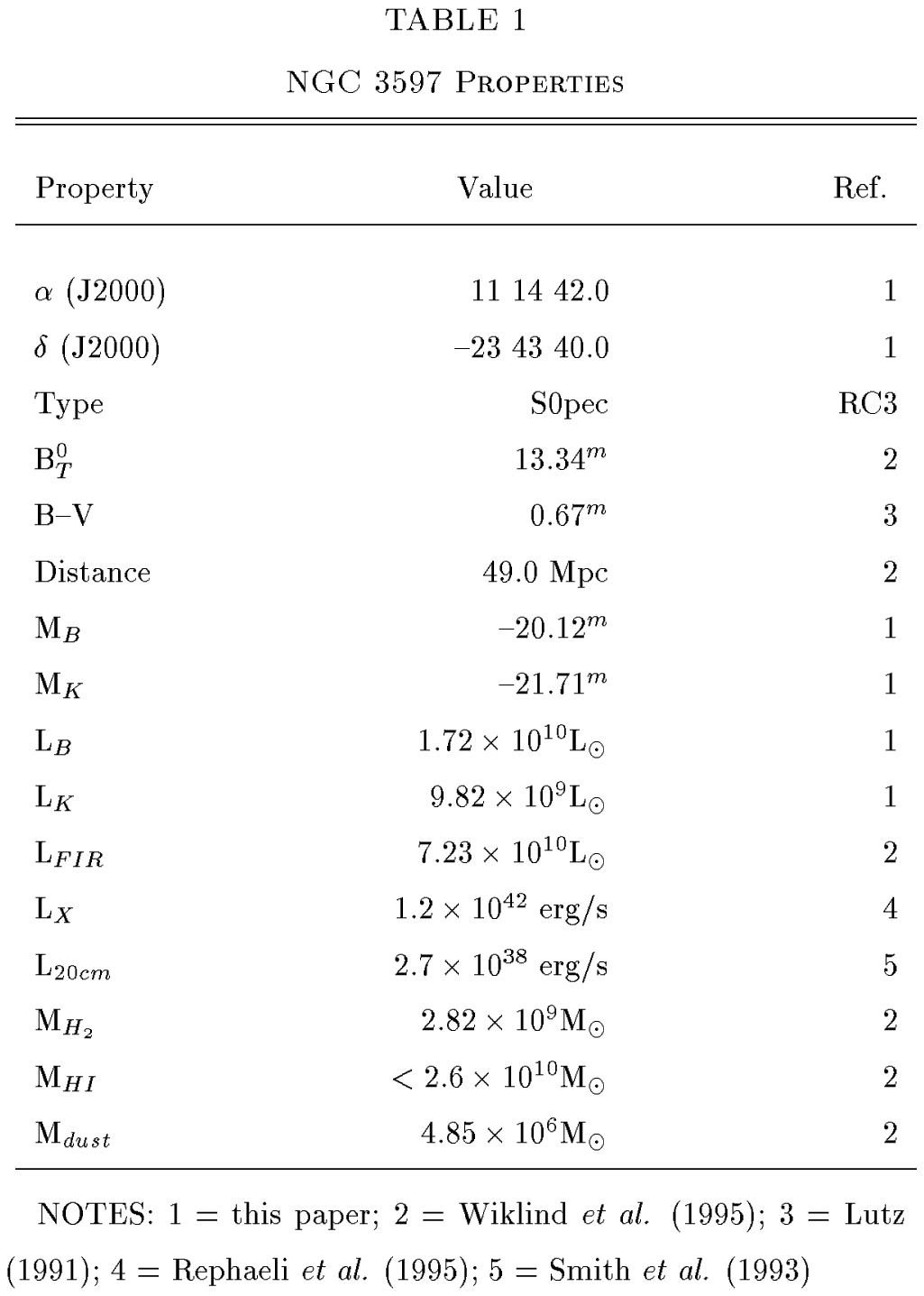,width=6.5in,height=6.5in}}
\end{figure*}

\begin{figure*}[p] 
\centerline{\psfig{figure=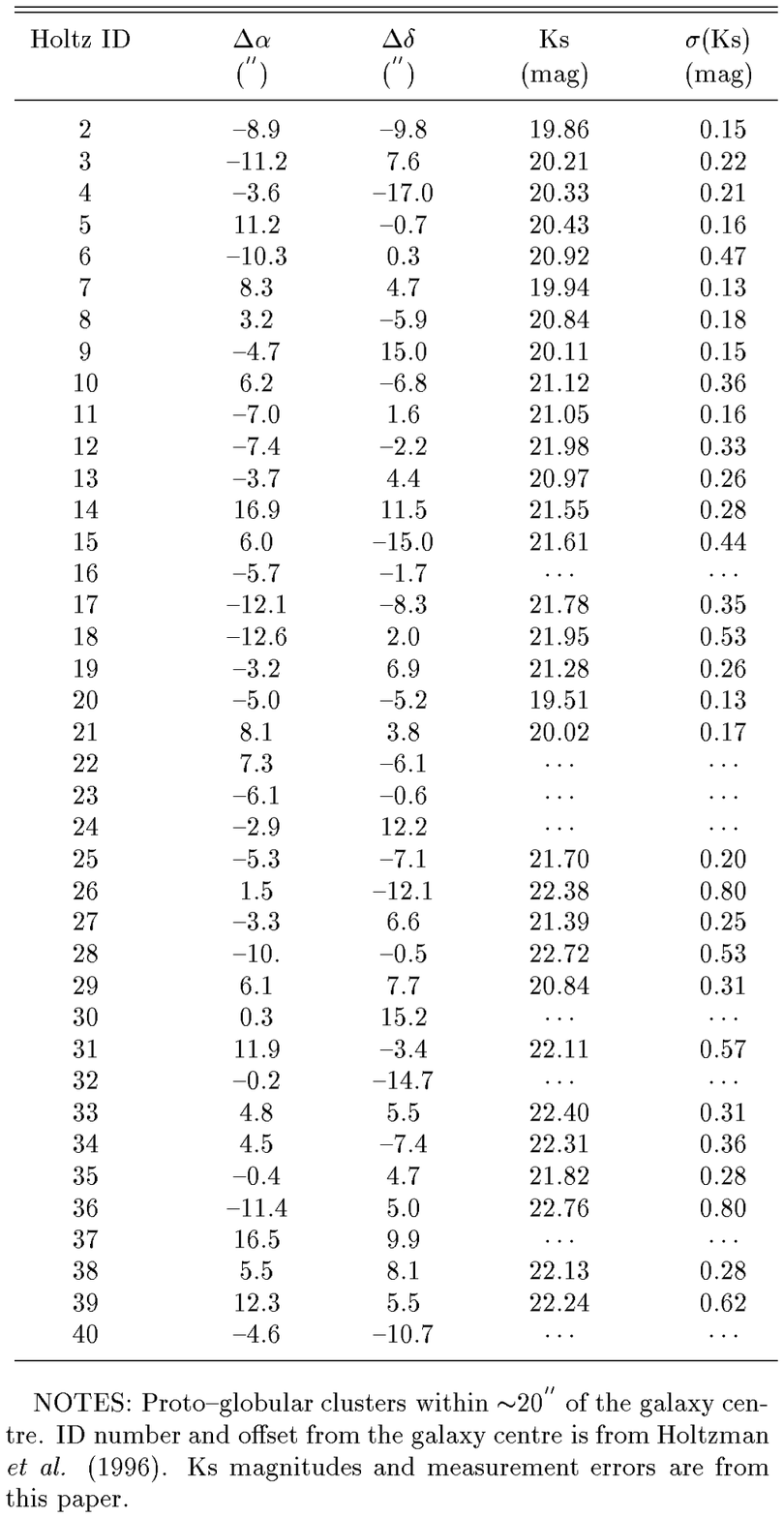,width=6.5in,height=6.5in}}
\end{figure*}

\begin{figure*}[p] 
\centerline{\psfig{figure=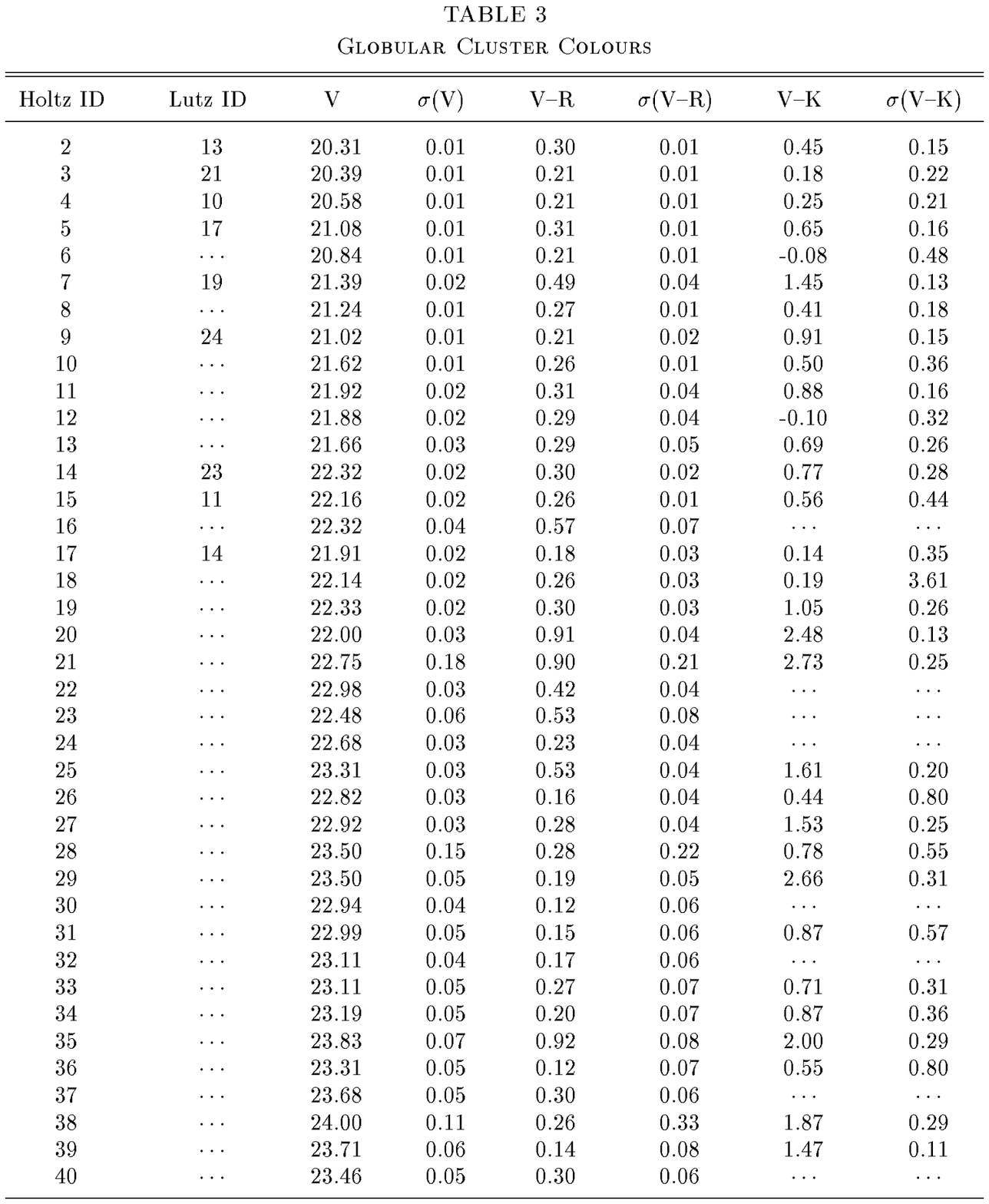,width=6.5in,height=6.5in}}
\end{figure*}

\begin{figure*}[p] 
\centerline{\psfig{figure=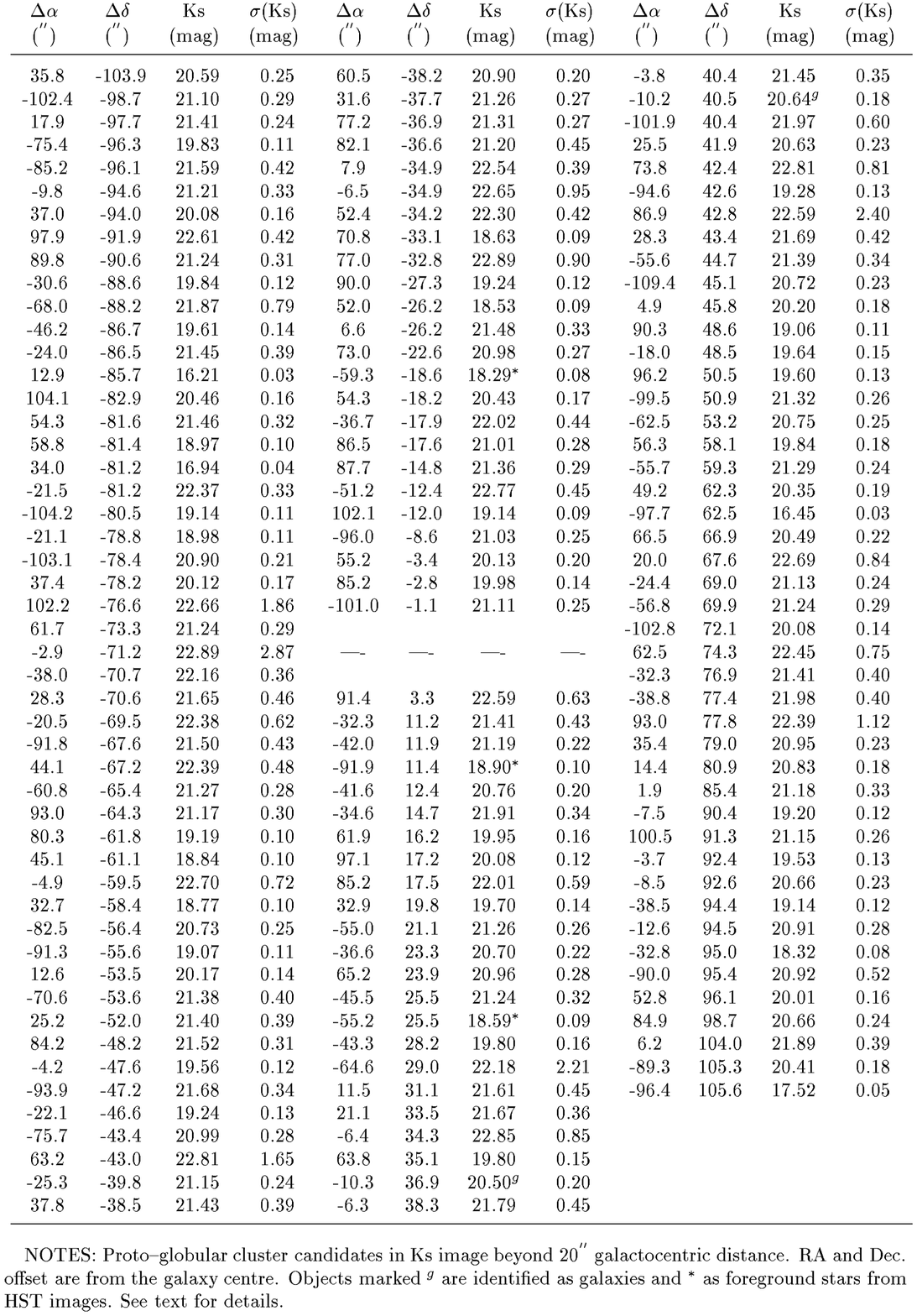,width=6.5in,height=6.5in}}
\end{figure*}

\end{document}